\documentclass[a4paper,11pt]{article}
\pdfoutput=1 

\usepackage{jheppub} 

\usepackage[T1]{fontenc} 
\usepackage{graphicx}
\usepackage{epstopdf}
\usepackage{epsfig}
\usepackage{amsmath}
\usepackage[caption=false]{subfig}
\usepackage{color}
\usepackage{amssymb}   
\usepackage{amsfonts}  
\usepackage{amsbsy}    
\usepackage{pstricks}
\usepackage{slashed}
\usepackage{dcolumn}
\usepackage{xparse}
\usepackage{bm}
\usepackage{hyperref}
\usepackage{soul}
\newcommand{\de}{\delta}

\newcommand{\la}{\lambda}

\newcommand{\beq}{\begin{equation}}
\newcommand{\eeq}{\end{equation}}
\newcommand{\ba}{\begin{array}}
\newcommand{\ea}{\end{array}}
\newcommand{\bea}{\begin{align}}
\newcommand{\eea}{\end{align}}
\newcommand{\bi}{\begin{itemize}}
\newcommand{\ei}{\end{itemize}}
\newcommand{\ben}{\begin{enumerate}}
\newcommand{\een}{\end{enumerate}}
\newcommand{\bc}{\begin{center}}
\newcommand{\ec}{\end{center}}
\newcommand{\bl}{\begin{flushleft}}
\newcommand{\el}{\end{flushleft}}
\newcommand{\br}{\begin{flushright}}
\newcommand{\er}{\end{flushright}}

\newcommand{\mc}{\mathcal}

\renewcommand{\Re}{{\mathrm{Re}}\,}
\renewcommand{\Im}{{\mathrm{Im}}\,}

\renewcommand{\l}{\left}
\renewcommand{\r}{\right}

\newcommand\comment[1]{ \hbox{[{\it Comment suppressed here.}\/]} }
\newcommand\hide[1]{}
\newcommand{\skipover}[1]{}

\newcommand{\cev}[1]{\reflectbox{\ensuremath{\vec{\reflectbox{\ensuremath{#1}}}}}}

\title{\boldmath Multiplicative noise and the diffusion
 of conserved densities}

\author[a,1]{Jingyi~Chao,\note{Corresponding author.}}
\author[b,2]{and Thomas~Sch\"afer \note{Corresponding author.}}

\affiliation[a]{Institute of Modern Physics, 
  Chinese Academy of Sciences,\\ Lanzhou, 730000, China}
\affiliation[b]{Department of Physics, North Carolina State University,\\
Raleigh, NC 27695}

\emailAdd{jychao@impcas.ac.cn}
\emailAdd{tmschaef@ncsu.edu}

\abstract{Stochastic fluid dynamics governs the long time tails of 
hydrodynamic correlation functions, and the critical slowing
down of relaxation phenomena in the vicinity of a critical point
in the phase diagram. In this work we study the role of multiplicative
noise in stochastic fluid dynamics. Multiplicative noise arises
from the dependence of transport coefficients, such as the 
diffusion constants for charge and momentum, on fluctuating
hydrodynamic variables. We study long time tails and relaxation
in the diffusion of a conserved density (model B), and a conserved
density coupled to the transverse momentum density (model H). 
Careful attention is paid to fluctuation-dissipation relations. 
We observe that multiplicative noise contributes at the same 
order as non-linear interactions in model B, but is a higher 
order correction to the relaxation of a scalar density and 
the tail of the stress tensor correlation function  in model H.}

\begin{document} 
\maketitle
\flushbottom

\section{Introduction}
\label{sec:intro}

 There is a very successful description of heavy ion collisions
at the Relativistic Heavy Ion Collider (RHIC) and the Large Hadron 
Collider (LHC) based on relativistic fluid dynamics
\cite{Romatschke:2017ejr,Jeon:2015dfa,Teaney:2009qa}. Up-to-date
models include higher order gradient corrections to the relativistic
Navier-Stokes theory, kinetic theory afterburners, and initial state
models that account for event-by-event fluctuations. 

 Recently, a number of authors have reexamined the role of fluctuations
in relativistic and non-relativistic fluid dynamics \cite{Schepper:1974,
Pomeau:1975,Kovtun:2003vj,PeraltaRamos:2011es,Kovtun:2011np,
Kapusta:2011gt,Kovtun:2012rj,Chafin:2012eq,Akamatsu:2016llw,
Martinez:2018wia,Chen-Lin:2018kfl,Akamatsu:2018vjr,An:2019osr}.
Fluctuations arise from the fact that fluid dynamics is a coarse-grained
description, and that the macroscopic variables arise from averaging 
over unresolved degrees of freedom at some resolution scale $l$. In 
approximate local thermal equilibrium these microscopic degrees of 
freedom exhibit thermal fluctuations that scale as $l^{-d/2}$,
where the $d$ is the number of spatial dimensions. As the resolution
scale becomes finer, the relative importance of fluctuations becomes
larger. Fluid dynamics is a non-linear theory, and the couplings between 
hydrodynamic modes lead to novel effects that go beyond Gaussian
noise in the macroscopic variables. A well-known example is the 
emergence of hydrodynamic "tails", non-analytic terms in the frequency
or time-dependence of correlation functions.

In this paper we will address a specific aspect of hydrodynamic
fluctuations, the role of non-linear or ``multiplicative'' noise.
Non-linear noise terms arise naturally in applications of fluid 
dynamics to relativistic heavy ion collisions. Fluctuation-dissipation
relations imply that the strength of noise terms is governed by 
dissipative coefficients, such as diffusion constants and viscosities. 
These coefficients are themselves functions of fluctuating hydrodynamic
variables, such as the entropy and baryon density of the fluid. As 
a consequence, noise terms are necessarily non-linear.

 There are a number of questions that immediately arise.  The first 
is how multiplicative noise fits into the power counting governed by 
the low energy (gradient) expansion. Naively, non-linear terms in the 
noise are not suppressed by extra gradients, so they might modify 
leading order predictions for the non-analyticities in correlation 
functions, or for the scaling behavior in the vicinity of a critical 
point. A second problem is to determine the precise form of the 
fluctuation-dissipation (FD) relation in the presence of multiplicative
noise. This problem has been studied in the past \cite{Wang:1998wg},
but the FD relations have not been checked in specific hydrodynamics 
theories of multiplicative noise. We also note that there is a substantial 
body of literature on stochastic equations with multiplicative noise
\cite{Habib:1993sf,Arnold:1999va,Arnold:1999uza,Aron:2010ac,Arenas:2012xy,
Biro:2004qg}.

   In this work we focus on the effect of multiplicative noise on 
the low energy expansion of hydrodynamic correlation functions. This 
paper is structured as follows: In Section \ref{sec-modB} we introduce 
an effective theory  of non-linear diffusion with multiplicative noise.
In Section \ref{sec-modB-cor} we formulate and check the FD relation, 
and compute corrections to the two-point function of the conserved
density. In Section \ref{sec-modH}-\ref{sec-modH-pipi} we extend this
analysis to the theory of a conserved density interacting with 
transverse shear waves (model H in the classification of Hohenberg 
and Halperin \cite{Hohenberg:1977ym}). We compute both the density 
and momentum density correlation functions, related to the relaxation
rate and the renormalization of the shear viscosity. 

\section{Diffusion}
\label{sec-modB}

 In this Section we study the diffusion of a conserved density 
$\psi(x,t)$. The diffusion equation is given by 
\beq 
\label{eq:modB}
\partial_t \psi(x,t)  = 
\vec\nabla \left\{ \kappa(\psi)\, \vec\nabla \,  
   \left(\frac{\delta {\cal F}[\psi]}{\delta\psi}\right) \right\} 
     + \theta(x,t) \, , 
\eeq
where $\kappa(\psi)$ is a density dependent conductivity, ${\cal F}
[\psi]$ is a free energy functional, and $\theta(x,t)$ is a noise term.
Equ.~(\ref{eq:modB}) is the diffusion equation of model B, modified 
by a field dependent conductivity $\kappa(\psi)$. In the following we 
will write 
\beq
 \kappa(\psi) = \kappa_0\left( 1 + \lambda_D \psi \right)\, , 
\eeq
and we will use a free energy functional of the form
\beq 
\label{F-modB}
{\cal F}[\psi]  = \int d^3x \, \left\{ 
 \frac{1}{2} (\vec\nabla \psi)^2
 +  \frac{r}{2}\, \psi(x,t)^2 +\frac{\lambda}{3!}\, \psi(x,t)^3 
   +h(x,t)\psi(x,t) \right\} \, ,
\eeq
where $h(x,t)$ is an external field. Higher order terms in $\kappa(\psi)$ 
and ${\cal F}[\psi]$ can be taken into account, but do not change our 
conclusions. The noise term $\theta(x,t)$ is Gaussian, with a distribution
\beq
\label{modB-noise}
P[\theta]\sim 
  \exp\left( -\frac{1}{4} \int d^3x\, dt\, \theta(x,t) L(\psi)^{-1}
     \theta(x,t)\right)\,  , 
\eeq
where $L$ is a noise kernel that we will specify below. Correlation
functions of this theory are computed from solutions of the diffusion
equation, averaged over the noise distribution in equ.~(\ref{modB-noise}).
Martin, Siggia, Rose, as well as Janssen and de Dominicis (MSRJD), showed 
how to write this noise average in terms of a stochastic field theory
\cite{Martin:1973zz,Janssen:1976,DeDominicis:1977fw}. This theory contains 
the hydrodynamic variable $\psi(x,t)$ as well as an auxiliary field
$\tilde\psi(x,t)$. The partition function is 
\beq
\label{Z-modB}
 Z = \int {\cal D}\psi\, {\cal D}\tilde\psi\;
  \exp\left(-\int d^3x\, dt\, {\cal L}(\psi,\tilde\psi)\right)\, . 
\eeq
The effective Lagrangian of this theory is 
\beq
\label{L-modB}
   {\cal L}(\psi,\tilde\psi) = 
     \tilde\psi \left( \partial_t -D_0 \nabla^2\right) \psi
     -\frac{D_0\lambda'}{2}\left(\nabla^2\tilde\psi\right) \psi^2
     -\tilde\psi L(\psi)\tilde\psi \, ,
\eeq
where we have defined the diffusion constant $D_0=r\kappa_0$ and 
$\lambda' = \lambda/r+\lambda_D$. We have dropped an $O(\nabla^4)$
term in the quadratic part of the Lagrangian, which is important in the
vicinity of a critical point when $r\to 0$. Note that in deriving this
Lagrangian we have dropped a Jacobian that can be written in terms of 
a set of ghost fields. As explained in \cite{Tauber:2014,Kovtun:2012rj}
ghost loops cancel pure vacuum diagrams that arise in the perturbative
expansion. In the following we do not explicitly write down ghost 
propagators and vertices, and simply drop pure vacuum diagrams. 

 An important observation is that for a suitable choice of the noise 
kernel $L(\psi)$ the effective Lagrangian enjoys a time reversal
symmetry \cite{Bausch:1976,Janssen:1979}. In the following we will 
choose
\beq
\label{modB-noise-ker}
 L(\psi) = \cev{\nabla}\l[k_BT \kappa(\psi)\r] \vec{\nabla}\, . 
\eeq
We will also employ units such that $k_BT=1$. We define the
$\mc{T}$-reversal of the stochastic fields as 
\begin{eqnarray}
\label{T-rev-psi}
  \mc{T} \psi(t)        &=&   \psi(-t) ,\\
\label{T-rev-psit}
  \mc{T} \tilde{\psi}(t)&=&  -\left(\tilde{\psi}(-t)
  - \frac{\delta{\cal F}[\psi(-t),\tilde\psi(-t)]}
     {\delta\psi(-t)}\right)\, . 
\end{eqnarray}
Under $\mc{T}$ the Lagrangian transforms as
\beq
\mc{T}{\cal L}(\psi(t),\tilde\psi(t)) = 
  {\cal L}(\psi(-t),\tilde\psi(-t)) - \frac{d}{dt}\,
    {\cal F}[\psi(-t),\tilde\psi(-t)]\, . 
\eeq
The total derivative term implies the detailed balance 
condition
\beq 
\exp\left(-\left[S(T_1,T_2)-S(-T_1,-T_2)\right]\right)
 = \exp\left(-\Delta {\cal F}\right) \, , 
\eeq
where 
\beq
S(T_1,T_2)=\int_{T_1}^{T_2} dt\, d^3x\,{\cal L}(\psi,\tilde\psi)\, ,
\eeq
and $\Delta{\cal F} = {\cal F}[\psi(T_2),\tilde\psi(T_2)] - 
{\cal F}[\psi(T_1),\tilde\psi(T_1)]$. The $\mc{T}$ invariance
of the MSRJD effective action was first studied in \cite{Janssen:1979}, 
for the case of a density-independent diffusion constant. We have 
verified that the symmetry continues to hold in the density 
dependent case, provided we use the symmetric noise kernel in 
equ.~(\ref{modB-noise-ker}). We note that the $T$-reversal 
symmetry in equ.~(\ref{T-rev-psi},\ref{T-rev-psit}) is closely
related to the dynamical KMS-symmetry studied in 
\cite{Crossley:2015evo,Glorioso:2017fpd,Chen-Lin:2018kfl}.

 Time reversal invariance can be used to derive fluctuation-dissipation
relations. For this purpose we define the response function as the 
derivative of $\langle \psi(t)\rangle_h$ with respect to the external 
field in equ.~(\ref{F-modB}). The response function is given by 
\beq
\label{modB-resp}
  G_R(x,t;x',t') = 
    \left.\frac{\delta \langle\psi(x,t)\rangle}
    {\delta h(x',t')}\right|_{h=0} \,  . 
\eeq
As explained in the appendix we can show that the response function is related to the correlation
function
\beq
\label{FD-modB}
\left\langle \psi(x_1,t_1)\l[\cev{\nabla} \kappa(\psi) \vec{\nabla}
  \tilde\psi\r](x_2,t_2)\right\rangle 
 = \Theta(t_2-t_1) \left\langle \psi(x_1,t_1)\dot\psi(x_2,t_2) 
 \right\rangle \, . 
\eeq  
This relation generalizes to higher order $n$-point functions. The
response of the $(n-1)$-point function is related to time-ordered 
$n$-point function. In momentum space equ.~(\ref{FD-modB}) is 
equivalent to 
\beq
\label{FD-modB-2}
  2\, {\rm Im}\;\left\{ k^2 
   \left\langle \psi(\omega,k) [\kappa(\psi) \tilde\psi](-\omega,-k) 
   \right\rangle\right\}
   = \omega \left\langle \psi(\omega,k) \psi(-\omega,-k) \right\rangle
   \, . 
\eeq
This is the standard FD relation in the case $\kappa(\psi)=\kappa_0$, 
but for a field dependent diffusion constant the left hand side of
equ.~(\ref{FD-modB-2}) includes the vertex function of the 
composite operator $[\kappa(\psi)\tilde\psi]$.

\section{Response and correlation functions}
\label{sec-modB-cor}

\begin{figure*}[t]
\begin{center}
\subfloat[]{%
    \includegraphics[width=8cm]{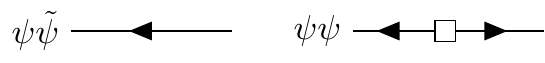}}\\
\subfloat[]{%
    \includegraphics[width=3.cm]{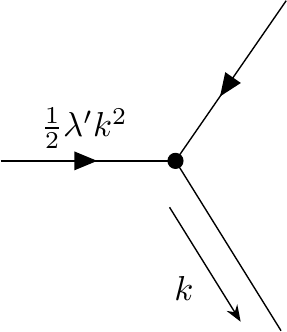}}
\hspace*{3.5cm}    
\subfloat[]{%
    \includegraphics[width=3.cm]{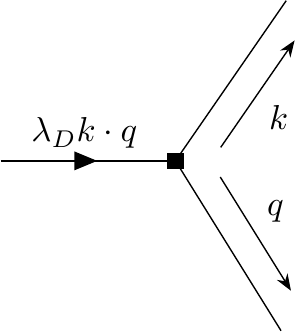}}
\end{center}
\caption{Feynman rules of the diffusive field theory (model B). The
vertex denoted by the solid circle is a non-linear interaction term, 
and the vertex denoted by a solid square is a new interaction generated
by the field dependence of the diffusion constant.
\label{fig:fey}}
\end{figure*}

  In this Section we will compute the response and correlation
functions of the purely diffusive theory at leading order in 
low frequency, low momentum expansion. For this purpose we split 
the effective Lagrangian into a quadratic and an interaction
part. The quadratic part of the action generates a matrix propagator 
in the $(\psi,\tilde\psi)$ basis. The off-diagonal matrix 
elements are retarded/advanced functions
\beq 
\label{modB-G0}
G_0(\omega,k) = \langle \tilde\psi \psi\rangle_{\omega,k} = 
\langle \psi \tilde\psi\rangle_{-\omega,k} = 
  \frac{1}{-i\omega+D_0k^2}\, , 
\eeq
and the diagonal components are the correlation function
\beq 
\label{modB-C0}
C_0(\omega,k)=\langle \psi \psi\rangle_{\omega,k} 
 = \frac{2\kappa_0k^2}{\omega^2+(D_0k^2)^2}\, , 
\eeq
as well as $\langle \tilde\psi \tilde\psi\rangle_{\omega,k}=0$.
The interaction term is 
\beq
\label{modB-int}
{\cal L}_{\it int} =  -\frac{D_0\lambda'}{2}
  \left(\nabla^2\tilde\psi\right) \psi^2
- \frac{D_0\lambda_D}{r} \left(\vec\nabla \tilde\psi\right)^2\psi\, ,
\eeq
and the corresponding vertices are shown in Fig.~\ref{fig:fey},
where we have set $r=1$.
We observe that both interaction terms involve two derivatives, 
and we expect the non-linear interaction and the field dependent
diffusion constant to contribute at the same order in the low 
energy expansion. However, the field dependent diffusion constant
leads to a new type of vertex not present in the standard 
MSRJD effective action. This type of vertex was previously obtained 
in \cite{Chen-Lin:2018kfl}, based on diffeomorphism invariance of 
the effective action on the Keldysh contour. 

\begin{figure}[t]
\begin{center}
\subfloat[]{%
    \includegraphics[width=3.5cm]{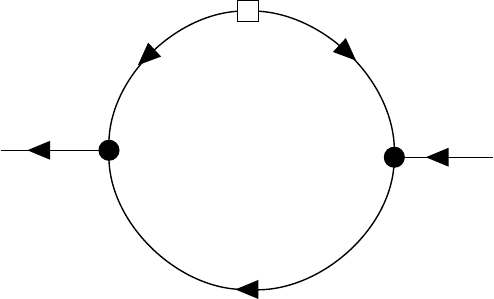}}
\hspace*{0.5cm}
\subfloat[]{%
    \includegraphics[width=3.5cm]{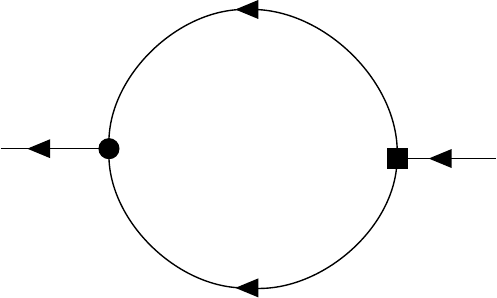}}
\hspace*{0.5cm}
\subfloat[]{%
    \includegraphics[width=3.5cm]{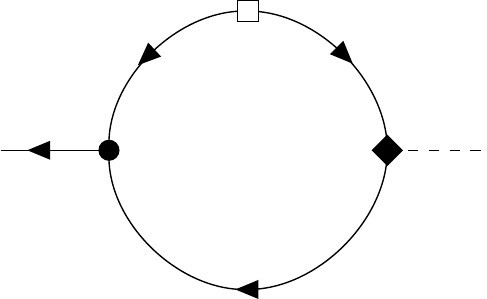}}
    \\
\subfloat[]{%
    \includegraphics[width=3.5cm]{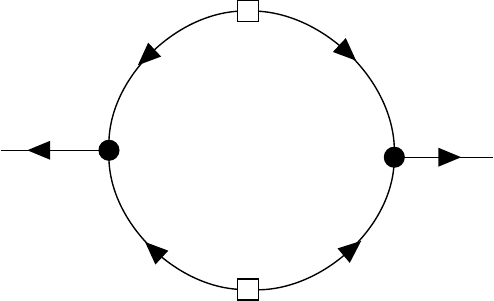}}
\hspace*{0.5cm}
\subfloat[]{%
    \includegraphics[width=3.5cm]{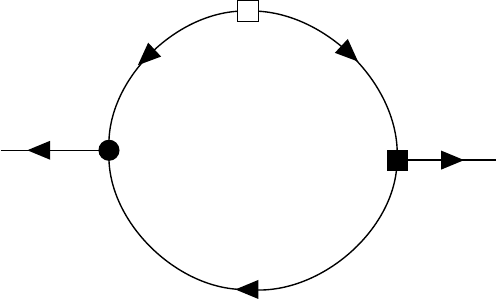}}
\end{center}
\caption{One-loop contributions to the response function (a)-(c) 
and the correlation function (d)-(e) in model B. Diagram (a) contains
only the non-linear interaction, whereas diagram (b) also depends
a new vertex generated by the field dependent diffusion constant.
Diagram (c) is a new type of diagram that contains the composite 
operator $\lambda_D [\psi\vec\nabla\tilde\psi]$. Diagrams (d,e) 
are the corresponding corrections to the correlation function.
\label{fig:Denn}}
\end{figure}

 One-loop corrections to the retarded and symmetric correlation
are shown in Fig.~\ref{fig:Denn}. Diagrams (a,b) are self energy
corrections to the retarded correlation functions. The self energy
modifies the retarded function as 
\beq 
\label{modB-G}
G(\omega,k) =   \frac{1}{-i\omega+D_0k^2+\Sigma(\omega,k)}\, . 
\eeq
At one-loop order, we find
\beq
\label{modB-Sig}
\Sigma(\omega,k) = \frac{\lambda'}{32\pi} 
 \left(   i\lambda'\omega k^2
   + \lambda_D \left[ i\omega -D_0 k^2\right]k^2
\right)
\sqrt{k^{2}-\frac{2i\omega}{D_0}} \, . 
\eeq
Here, we have regularized the loop integral and dropped 
cutoff-dependent terms that can be absorbed into the bare 
diffusion constant. The result in equ.~(\ref{modB-Sig}) shows 
that $\lambda_D$ indeed contributes at the same order as ordinary
non-linear interactions. We note, however, that the functional 
form of the correction is different, so that it is possible to
disentangle the corrections from $\lambda'$ and $\lambda_D$. 
Finally, we note that in the limit $k^2\to 0$ the coefficient 
of the self energy is shifted $\lambda'^2\to \lambda'(\lambda' +
\lambda_D)$, indicating that the density dependence of $D_0$ 
corrects the long-time tail of the response function by an overall 
factor $(1+\lambda_D/\lambda')$.

Diagrams (d,e) provide corrections to the correlation function,
which is modified as 
\beq 
\label{modB-C}
C(\omega,k) =   
\frac{ 2D_0k^2+\delta D(\omega,k)}
{\left(-i\omega+D_0k^2+\Sigma(\omega,k)\right)
 \left(i\omega+D_0k^2+\Sigma(-\omega,k)\right)}  \,  ,
\eeq
where 
\beq
\delta D(\omega,k) = \frac{D_0\lambda'}{16\pi}
 \left( \lambda' + 2\lambda_D \right) k^4\; 
\Re\, \sqrt{k^{2}-\frac{2i\omega}{D_0}} \, . 
\eeq
Diagram (c) shows the one-loop contribution with one insertion
of the composite operator $\lambda_D[\psi\vec\nabla\tilde\psi]$. 
We define the corresponding vertex function by
\beq
\Gamma_D(\omega,k) \equiv  (-i\omega+D_0 k^2) 
  \left\langle D_0\lambda_D[\psi\vec\nabla\tilde\psi] \vec\nabla\psi
   \right\rangle_{\omega,k}  
\eeq 
and get 
\beq
\label{modB-noise-vert}
\Gamma_D(\omega,k) = 
\frac{D_0\lambda'\lambda_D}{32\pi} \, k^4 \, 
\sqrt{k^{2}-\frac{2i\omega}{D_0}}\, . 
\eeq
We can now verify the fluctuation-dissipation relation. In terms of
the Green and vertex functions defined in this Section the FD
relation in equ.~(\ref{FD-modB-2}) becomes
\beq
\label{FD-modB-3}
 2\, \Im \left\{  G(\omega,k) \left[ D_0k^2 + 
   \Gamma_D(\omega,k) \right] \right\}
    = \omega C(\omega,k)\, . 
\eeq
Using equ.~(\ref{modB-G}-\ref{modB-noise-vert}) we observe that 
this relation is indeed satisfied. In the limit $\lambda_D=0$
(no multiplicative noise) equ.~(\ref{FD-modB-3}) reduces to the
well known relation between the retarded function and the correlation
function. However, in the presence of multiplicative noise the 
contribution of the vertex function $\Gamma_D$ is essential in 
satisfying the FD relation. 

\section{Shear modes and model H}
\label{sec-modH}

 In this Section we will extend our result to a conserved 
density $\psi$ that is advected by the momentum density $\vec\pi$ 
of a fluid. This theory is known as model H \cite{Hohenberg:1977ym},
and describes the critical behavior of a fluid near the liquid-gas
endpoint. In the following we will assume that the fluid is 
approximately incompressible, $\nabla_k\pi_k\simeq 0$. This implies
that we are only taking into account the coupling to shear modes, 
neglecting the role of sound. This approximation is sufficient 
to capture the critical dynamics in model H, and to compute 
the shear contribution to hydrodynamic tails. 

\begin{figure*}[t]
\begin{center}
\subfloat[]{%
    \includegraphics[width=8cm]{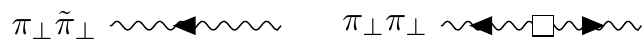}}\\
\subfloat[]{%
    \includegraphics[width=3.cm]{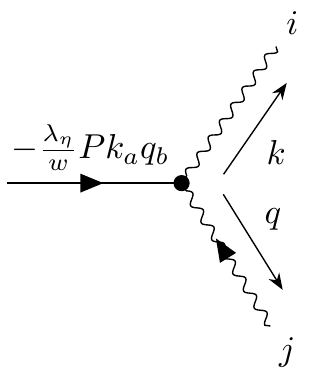}}
\hspace*{1.cm}    
\subfloat[]{%
    \includegraphics[width=3.cm]{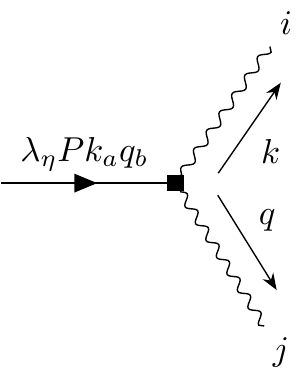}}
\hspace*{1.cm}    
\subfloat[]{%
    \includegraphics[width=3.cm]{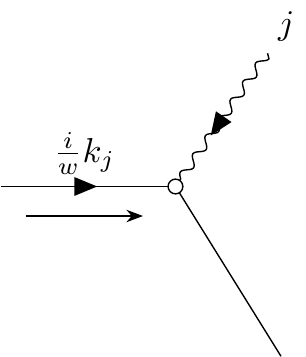}}
\hspace*{1.cm}   
\subfloat[]{%
    \includegraphics[width=3.cm]{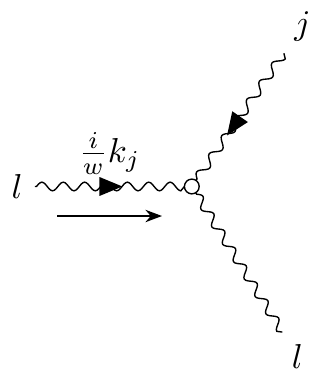}}
\end{center}
\caption{Feynman rules for model H, describing the interaction
of the transverse momentum density $\pi_k$ with a scalar density
$\psi$. The vertex denoted by the solid circle is a non-linear
interaction term, and the vertex denoted by a solid square is a 
new interaction generated by the field dependence of the diffusion
constant. Note: $P_{abij}=\de_{ab}\de_{ij}$.
\label{fig:H:fey}}
\end{figure*}

In the presence of a conserved momentum density the diffusion
equation contains a new coupling, $\partial_t \psi(x,t) \sim w^{-1} 
\pi_k\nabla_k \psi(x,t)$, where $w$ is the enthalpy density of the 
fluid. In order to obtain the correct equilibrium distribution in 
the presence of this coupling it is important to note that this 
interaction derives from a Poisson bracket \cite{Dzyaloshinski:1980}
\beq
\label{modB-PB}
\partial_t \psi(x,t) = \ldots - \int d^3x'\,
 \l\{\psi(x,t),\pi_{k}(x',t\r\} \frac{\de \mc{F}[\psi,\pi_l]}{\de\pi_{k}}
  = \ldots - \frac{1}{w} \pi_k\nabla_k \psi(x,t)\, , 
\eeq
where $\ldots$ denotes the right hand side of equ.~(\ref{eq:modB})
and the free energy density is 
\beq
\label{F-pi}
\mc{F}[\psi,\pi_k] = \mc{F}[\psi] + 
   \int d^3x\, \left\{ \frac{1}{2w}\, \vec\pi^2 
   + \pi_k {\cal A}_k \right\}\, , 
\eeq
where $\mc{F}[\psi]$ is defined in equ.~(\ref{F-modB}) and ${\cal A}_k$
is an external field coupled to the momentum density. Note that in a fully
covariant formalism ${\cal A}_k$ corresponds to the $g_{0k}$ components 
of the metric tensor. The equation of motion for the momentum density is 
\beq
\label{modH-PB}
\partial_t \pi_i(x,t) =  \vec\nabla  \left\{ 
  \eta(\psi)\, \vec\nabla \,  
   \left(\frac{\delta {\cal F}[\psi,\pi_k]}{\delta\pi_i}\right) \right\}  
    + \frac{\delta \mc{F}[\psi,\pi_k]}{\delta\psi}\nabla_i\psi
    - \frac{\delta \mc{F}[\psi,\pi_l]}{\delta\pi_k} \nabla_k \pi_i 
     + \xi_i(x,t) \, ,
\eeq
where we have neglected terms proportional to $\nabla_k\pi_k$ and
$\xi_i$ is a stochastic force. The stochastic force has a Gaussian
probability distribution
\beq
\label{modH-noise}
P[\xi_i]\sim 
  \exp\left( -\frac{1}{4} \int d^3x\, dt\, \xi_i(x,t) 
    \l[ M(\psi)^{-1}\r]_{ij}
     \xi_j(x,t)\right)
\eeq
with 
\beq
\label{modH-noise-ker}
 M_{ij}(\psi) = \delta_{ij}\cev{\nabla} \eta(\psi) \vec{\nabla}\, .
\eeq
The noise kernel can be generalized for compressible fluids. In
the following we will only use that $M$ is symmetric. As before
we will take the dependence on $\psi$ to be linear
\beq 
\eta(\psi) = \eta_0 \l( 1 + \lambda_\eta\psi \r)\, . 
\eeq
The two Poisson bracket terms in equ.~(\ref{modH-PB}) can be written as
\beq
\label{modH-PB-2}
\partial_t \pi_i(x,t) = \ldots - \l(\nabla^2\psi\r)\nabla_i\psi - 
  \frac{1}{w}\pi_k\nabla_k \pi_i \, . 
\eeq
The quadratic part of the effective Lagrangian for the momentum density is 
\beq
\label{L-modH}
   {\cal L}(\pi_i,\tilde\pi_i) = 
     \tilde\pi_i \left( \partial_t - \gamma_0 \nabla^2\right) \pi_i
     -\tilde\pi_i M(\psi)_{ij}\tilde\pi_j \, ,
\eeq
where $\gamma_0=\eta_0/w$, and the fields are understood to 
satisfy $\nabla_k\pi_k=0$. The interaction term is 
\beq
\label{L-modH-int}
{\cal L}_I = \frac{1}{w}\tilde\psi\pi_k\nabla_k\psi
  + \tilde\pi_k\l(\nabla^2\psi\r)\nabla_k\psi
  + \frac{1}{w}\tilde\pi_i \pi_k\nabla_k\pi_i \, , 
\eeq
where the first term corresponds to advection of $\psi$ by 
$\pi_k$, the second term is a higher order correction that 
describes the coupling of $\pi_k$ to $\nabla_k\psi$, and the
third term is the advection of $\pi_k$ by the momentum itself. 
Multiplicative noise generates a noise vertex
\beq
{\cal L}_n = \gamma_0\lambda_\eta \psi \l\{ 
  \frac{1}{w}  \l(\nabla_i\tilde\pi_k\r) \l(\nabla_i\pi_k\r)
 -  \l(\nabla_i\tilde\pi_k\r) \l(\nabla_i\tilde\pi_k\r)
\r\} \, . 
\eeq
As in the case of model B, this effective Lagrangian is 
invariant under time reversal. The $\mc{T}$ transformation of the 
momentum density is 
\begin{eqnarray}
  \mc{T} \pi_k(t)       &=&   -\pi_k(-t) ,\\
  \mc{T} \tilde{\pi}_k(t)&=&  +\left(\tilde{\pi}_k(-t)
  - \left.\frac{\delta{\cal F}[\psi,\pi_i]}
     {\delta\pi_k}\right|_{\mc{T}[\psi,\pi]}\right)\, . 
\end{eqnarray}
We note that the intrinsic $\mc{T}$-parity of $\pi_k$ is negative. 
As before, the Lagrangian is invariant up to a total time derivative 
of the free energy density. In order to show the invariance of the 
Lagrangian we have to use three ingredients: 1) The dissipative 
matrix $M$ is symmetric, 2) the mode coupling matrix is 
anti-symmetric, and 3) the mode coupling matrix is 
$\mc{T}$-odd. The last two ingredients follow from the properties 
of Poisson brackets. Time reversal invariance can again be used
to derive a fluctuation-dissipation relation. The new ingredient
in the presence of mode couplings is a new response vertex induced 
by the Poisson bracket terms. Consider the variational derivative 
in equ.~(\ref{modB-resp}) acting on the Poisson bracket term in
equ.~(\ref{modH-PB}). This variation induces a new composite operator 
$X=\tilde{\pi}_{k}\nabla_{k}\psi$. The FD relation for the 
density-density correlation function is given by 
\beq
\label{FD-modH-1}
 2\, \Im \left\{  G(\omega,k) \left[ D_0k^2 + 
     \Gamma_D(\omega,k) + \Gamma_X(\omega,k)\right] \right\}
    = \omega C(\omega,k)\, ,
\eeq
where the vertex function $\Gamma_X$ is defined by 
\beq
\label{Gamma-X}
 \Gamma_{X}(\omega,k) = \l(-i\omega+D_0 k^2\r) 
  \left\langle \psi X\right\rangle_{\omega,k} \, .
\eeq

\section{Advection of the scalar density}
\label{sec-modH-nn}

\begin{figure*}[t]
  \centering
  \subfloat[]{%
    \includegraphics[width=3.5cm]{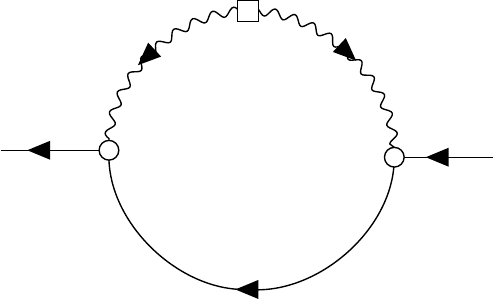}}
\hspace*{0.5cm}
\subfloat[]{%
    \includegraphics[width=3.5cm]{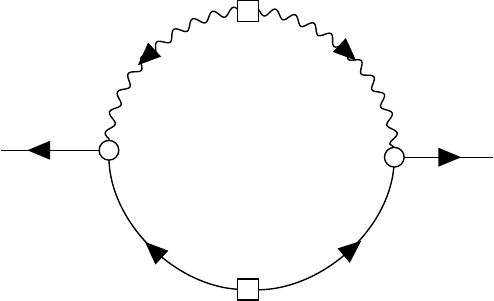}}
\\
  \subfloat[]{%
    \includegraphics[width=3.5cm]{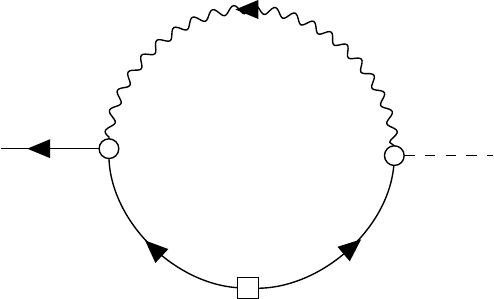}}
\caption{Contributions to the density response and density 
correlation that arise from the coupling to the momentum
density. Fig.~(a) and (b) show the retarded function and 
correlation function at leading order in the advective 
couplings. Fig.~(c) shows the vertex function for the composite 
operator $X$ that appears in the FD relation.
\label{fig:modH-nn}}
\end{figure*} 

  In this Section we consider corrections to the density response
induced by the coupling to the momentum density. These terms are 
of interest for two reasons: 1) As we will see, the coupling to 
$\pi_k$ generates the leading hydrodynamic tail in the density 
response, and 2) in a critical fluid the order parameter relaxation
rate is dominated by the coupling to the momentum density (the
corresponding momentum dependent relaxation rate is known as the 
Kawasaki function \cite{kawasaki:1970}). 
  
 Feynman diagrams for the leading order corrections to the response 
and correlation functions are shown in Fig.~\ref{fig:modH-nn}. 
The retarded functions for the momentum density is given by 
\beq 
\label{modH-G0}
G_{\pi,0}^{ij}(\omega,k) = \langle \tilde\pi^i \pi^j\rangle_{\omega,k} 
  =  \frac{P^{ij}_\perp(k)}{-i\omega+\gamma_0k^2}\, , 
  \hspace{0.5cm}
  P^{ij}_\perp(k)= \delta^{ij}-\hat{k}^i\hat{k}^j\, , 
\eeq
where $P^{ij}_\perp(k)$ is a transverse projection operator and 
$\hat{k}=\vec{k}/|\vec{k}|$ is a unit momentum vector. The 
correlation function is 
\beq 
\label{modH-C0}
C_{\pi,0}^{ij}(\omega,k)=\langle \pi^i \pi^j\rangle_{\omega,k} 
 = \frac{2\gamma_0 w k^2P_\perp^{ij}(k)}{\omega^2+(\gamma_0k^2)^2}
\eeq
and the interaction vertices are summarized in Fig.~\ref{fig:H:fey}.
Fig.~\ref{fig:modH-nn}(a) corresponds to a self energy term. We get
\beq
\label{modH-Sig-nn}
\Sigma(\omega,k) = \frac{1}{6\pi}
  \frac{k^2}{w(\gamma_0+D_0)}\sqrt{\frac{-i\omega}{\gamma_0+D_0}}\, , 
\eeq
where, for simplicity, we have expanded the result to leading order 
in $k^2$ for $\omega\neq 0$. We can reinstate the temperature by 
replacing $w\to w/T$. Note that this result is more important, 
in the sense of the gradient expansion, than the contribution from
non-linear interactions and multiplicative noise, see 
equ.~(\ref{modB-Sig}). Equ.~(\ref{modH-Sig-nn}) determines the leading
hydrodynamic tail in the density response, and it has been computed 
many times in the literature, see \cite{Schepper:1974,Kovtun:2003vj,
Kovtun:2012rj,Martinez:2018wia}. The fact that equ.~(\ref{modH-Sig-nn}) 
is lower order in $k^2$ compared to equ.~(\ref{modB-Sig}) can be traced 
to the fact that mode coupling vertices are $O(k)$, whereas the 
non-linear interaction and noise vertices are $O(k^2)$.

The one-loop correction to the correlation function is shown in 
Fig.~\ref{fig:modH-nn}(b). This diagram can be viewed as a contribution
to $\delta D$ in equ.~(\ref{modB-C}). We find
\beq
\label{modH-D-nn}
\delta D(\omega,k) = \frac{1}{3\pi}
  \frac{k^2}{w(\gamma_0+D_0)}
   \; \Re\, \sqrt{\frac{-i\omega}{\gamma_0+D_0}}\, .
\eeq
Equ.~(\ref{modH-Sig-nn}) and (\ref{modH-D-nn}) do not satisfy 
naive FD relation. Instead, we have to include the contribution 
of the vertex function $\Gamma_X$, given by 
\beq
\label{modH-X-nn}
\Gamma_X(\omega,k) = \frac{1}{6\pi}
  \frac{k^2}{w(\gamma_0+D_0)}\sqrt{\frac{-i\omega}{\gamma_0+D_0}}\, .
\eeq
We can now verify that the extended FD relation (\ref{FD-modH-1})
is satisfied. We also note that Figs.~\ref{fig:Denn} and \ref{fig:modH-nn}
comprise the full set of one-loop corrections to the density 
response in the presence of advection and multiplicative noise. 
We note, in particular that multiplicative noise does not modify 
the leading order result in equ.~(\ref{modH-Sig-nn}). This means 
that it does not modify the Kawasaki function, which is the self energy
$\Sigma(\omega,k)$ in the limit $\omega\to 0$ and $r\to 0$. The 
Kawasaki function governs the dynamical critical exponent $z\simeq 
3$ of model H \cite{Hohenberg:1977ym,Onuki:2002}.

\section{Renormalization of the shear viscosity}
\label{sec-modH-pipi}

\begin{figure}[t]
\begin{center}
\subfloat[]{%
    \label{fig:H:respcon}
    \includegraphics[width=3.5cm]{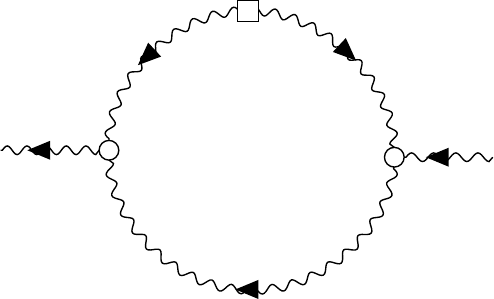}}
\hspace*{0.5cm}
\subfloat[]{%
    \label{fig:H:corrcon}
    \includegraphics[width=3.5cm]{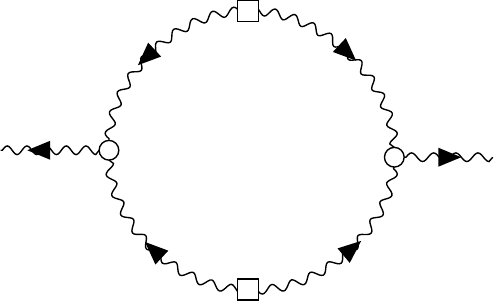}}
\hspace*{0.5cm}
\subfloat[]{%
    \label{fig:H:3ptcon}
    \includegraphics[width=3.5cm]{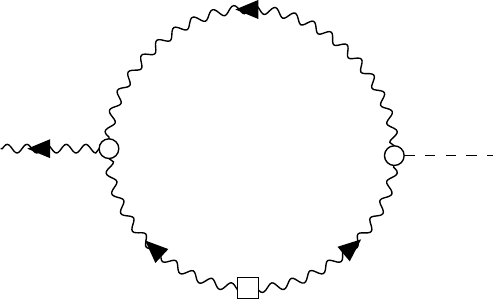}}
    \\
\subfloat[]{%
    \label{fig:H:resp1}
    \includegraphics[width=3.5cm]{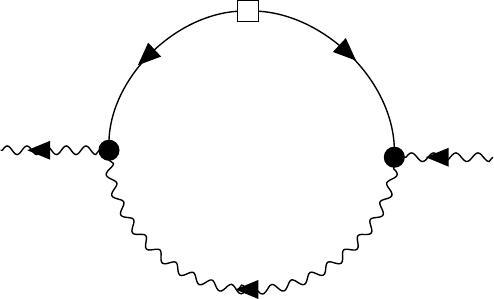}}
\hspace*{0.5cm}
\subfloat[]{%
    \label{fig:H:corr2}
    \includegraphics[width=3.5cm]{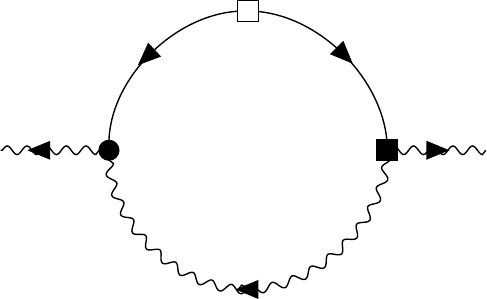}}
\hspace*{0.5cm}
\subfloat[]{%
    \label{fig:H:corr1}
    \includegraphics[width=3.5cm]{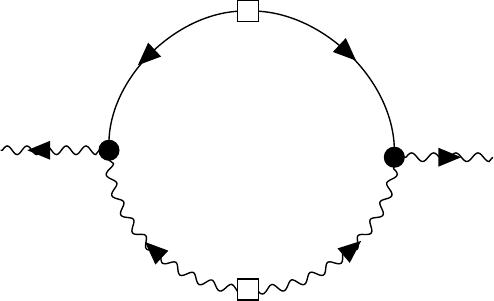}}
\\
\subfloat[]{%
    \label{fig:H:3pt}
    \includegraphics[width=3.5cm]{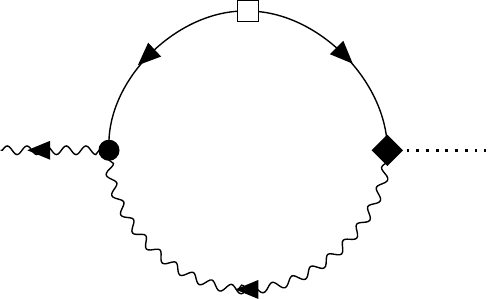}}
\end{center}
\caption{One-loop contributions to the response and correlation function 
in model H. Diagrams (a) and (b) contain the mode coupling interaction, 
diagram (c) is the vertex function that appears in the FD relation, 
and diagrams (d)-(g) are induced by multiplicative noise.  
\label{fig:modH-pipi}}
\end{figure}

 In this Section we study the response function of the transverse
momentum density. At tree level, this response is controlled by 
the momentum diffusion constant $\gamma_0=\eta/w$. The leading 
correction to the response arises from the diagram in 
Fig.~\ref{fig:modH-pipi}(a). This diagram corresponds to a self 
energy term. We define the transverse self energy  $\Sigma_T$ by 
$\Sigma^{ij} =\Sigma_T P_\perp^{ij}(k) +\Sigma_L \hat{k}^i\hat{k}^j$
and find
\beq
\label{modH-Sig-T}
\Sigma_T(\omega,k) = \frac{7}{120\pi}\frac{k^2}{w\gamma_0}
\sqrt{\frac{-i\omega}{2\gamma_0}}\, ,
\eeq
where we have expanded the result to leading order in $k^2$ for $\omega
\neq 0$, and we can reinstate the temperature by replacing $w\to w/T$.
As before, we do not explicitly write a frequency independent term that 
is linearly divergent in the cutoff $\Lambda$. This contribution can be 
viewed as a term that combines with the bare momentum diffusion constant
to provide the physical diffusivity. Equ.~(\ref{modH-Sig-T}) determines
the leading hydrodynamic tail in the stress tensor correlation function
in a theory with only shear modes. The numerical coefficient in 
equ.~({\ref{modH-Sig-T}) agrees with the result in \cite{Kovtun:2011np,
Chafin:2012eq,Akamatsu:2016llw,Martinez:2018wia}.
Fig.~\ref{fig:modH-pipi}(b) shows the corresponding contribution to 
the correlation function. In analogy with equ.~(\ref{modB-C}) we 
define a correction $\delta\gamma$ to the numerator of the correlation
function. We obtain
\beq
\label{modH-gam-T}
\delta\gamma_T(\omega,k) = \frac{7}{60\pi}\frac{k^2}{w\gamma_0}
\, \Re\,\sqrt{\frac{-i\omega}{2\gamma_0}}\, .
\eeq 
In order to satisfy the FD relation we have to include a new vertex 
function $\Gamma_Y$, where the composite operator $Y$ is given by $Y_k=
\tilde\pi_i\nabla_k\pi_i-\tilde\pi_i\nabla_i\pi_k$. The structure of
$Y_k$ follows from the symmetry of the Poisson bracket $\{\pi_i,\pi_k\}$.
The vertex function $\Gamma_Y$ is defined by  
\beq
 \Gamma_{Y,T}(\omega,k) = \l(-i\omega+\gamma_0 k^2\r) P_{jk}^\perp(k)
  \left\langle \pi_{j}Y_k\right\rangle_{\omega,k}
\eeq
Computing the diagram in Fig.~\ref{fig:modH-pipi}(c) we obtain
\beq
 \Gamma_{Y,T}(\omega,k) = \frac{7}{120\pi}\frac{k^2}{w\gamma_0}
\sqrt{\frac{-i\omega}{2\gamma_0}}\, .
\eeq
This result satisfies the FD relation
\beq
\label{FD-modH-2}
 2\, \Im \left\{  G_T(\omega,k) \left[ \gamma_0 wk^2 
  + \Gamma_{Y,T}(\omega,k)\right] \right\}
    = \omega C_T(\omega,k)\, ,
\eeq
where $G_T$ and $C_T$ denote the transverse retarded function
and correlation function, respectively. We note that there is 
a higher order correction to equ.~(\ref{modH-Sig-T},\ref{modH-gam-T}) 
that arises from the $O(\nabla^2)$ mode coupling to $\psi$ in
equ.~(\ref{L-modH-int}). We do not study this term here.

Fig.~\ref{fig:modH-pipi}(d) shows the leading correction to the 
transverse self energy due to multiplicative noise.
Fig.~\ref{fig:modH-pipi}(e,f,g) are the corresponding corrections
to the correlation and vertex function. The self energy is 
\beq
\label{modH-Sig-T-mult}
\Sigma_T(\omega,k) = \frac{(\gamma_0\la_{\eta})^2}{15\pi w}
  \; k^2 \; \left(\frac{-i\omega}{\gamma_0+D_0}\right)^{3/2}
  \, .
\eeq
Here, we have expanded $\Sigma_T(\omega,k)$ to leading order 
in $k^2$ for $\omega\neq 0$. We have also dropped cutoff dependent
terms that renormalize the transport coefficients. We observe
that multiplicative noise modifies the low frequency behavior
of the shear viscosity, but that this correction is subleading 
compared to equ.~(\ref{modH-Sig-T}). It is known that in model 
H the critical enhancement of the shear viscosity is very weak
\cite{Onuki:2002}. Our results indicate that this result is 
not modified by multiplicative noise. 

\section{Conclusions and outlook}
\label{sec-sum}

 In this work we studied the role of multiplicative noise in the 
theory of a conserved density coupled to the transverse momentum 
density of a fluid. This theory governs the critical behavior of 
both ordinary fluids \cite{Hohenberg:1977ym} and the quark gluon 
plasma \cite{Son:2004iv} in the vicinity of a possible critical end 
point. In this context we can think of $\psi$ as the entropy
per particle of the fluid. Multiplicative noise arises from the 
dependence of the thermal conductivity and shear viscosity 
on $\psi$. 

 Multiplicative noise is consistent with suitably generalized
fluctuation-dissipation relations. It also fits into the standard 
long time, large wavelength, expansion of hydrodynamic correlation 
functions. We find that multiplicative noise contributes to the 
long-time tails of the density and momentum density correlation
functions. In model B, without the coupling to the momentum
density, this contribution is leading order. In model H the 
multiplicative noise contribution to the tails is subleading 
compared to the contributions induced by mode couplings. At 
leading order in $k$ multiplicative noise does not modify the 
Kawasaki function, which governs the order parameter relaxation
rate in model H \cite{Hohenberg:1977ym}, or the self energy of
transverse momentum modes, which determines the renormalization
of the shear viscosity.

 In the present work we have used diagrammatic methods to study 
correlation functions in a fluid at rest, or in the local restframe
of a slowly evolving background flow. These results are applicable 
to both relativistic and non-relativistic systems. We have not 
addressed the issue of writing the effective action in a manifestly 
covariant way \cite{Kovtun:2014hpa}, or investigated correlation 
functions in an evolving background \cite{Akamatsu:2016llw,An:2019osr}. 
We have also not studied the coupling to sound modes, and the 
renormalization of the bulk viscosity in a non-conformal fluid 
\cite{Kovtun:2003vj,Martinez:2017jjf,Akamatsu:2017rdu,
Martinez:2019bsn}. Finally, it would be interesting to study 
multiplicative noise in numerical simulations of stochastic 
diffusion in an expanding fluid \cite{Nahrgang:2018afz}.

\acknowledgments

The work of T.~S.~was supported in part by the US Department of 
Energy grant DE-FG02-03ER41260 and by the BEST (Beam Energy Scan 
Theory) DOE  Topical Collaboration. J.~C. was supported by the 
Major State Basic Research Development Program in China 
(No.~2015CB856903).

\appendix
\section{Fluctuation-Dissipation relation}
\label{sec:FD}

 In this appendix we derive the fluctuation-dissipation (FD) 
relation in model B in the presence of multiplicative noise,
following the method described in \cite{Janssen:1979,Tauber:2014}.
For this purpose we compute the response function 
Equ.~(\ref{modB-resp}) in two different ways. The first 
is based on coupling an external source to the MSRJD 
functional. The source term in the effective Lagrangian
equ.~(\ref{L-modB}) is ${\cal L}_h = \tilde\psi(x,t)L(\psi)h(x,t)$,
where $L(\psi)$ is defined in equ.~(\ref{modB-noise-ker}). We
can then compute the response function 
\beq
\label{GR_MSRJD}
 G_R(x_1,t_1;x_2,t_2) =  \left\langle 
\psi(x_1,t_1)\l[ L(\psi)\tilde\psi\r](x_2,t_2)\right\rangle \, . 
\eeq
The second method is based on the noise average in 
equ.~(\ref{modB-noise}). Formally, we can express the noise
in terms of the diffusion equation, 
\beq
\theta(x,t)  = \partial_t\psi(x,t)+L(\psi)
   \frac{\delta{\cal F}[\psi]}{\delta\psi}\, . 
\eeq
This implies that we perform the average over solutions 
of the diffusion equation with respect to the noise
distribution in equ.~(\ref{modB-noise}) by changing 
variables from $\theta(x,t)$ to $\psi(x,t)$. The 
corresponding partition function is 
\beq 
\label{Z_OM}
  Z_{OM} = \int {\cal D}\psi\, \exp\l(-S_{OM}(\psi)\r)\, , 
\eeq
where $S_{OM}$, the Onsager-Machlup functional, is 
given by 
\beq
S_{OM} = -\frac{1}{4}\int dt\, dx\, 
 \l(\partial_t \psi +L(\psi)
   \frac{\delta{\cal F}[\psi]}{\delta\psi} \r)
  \l[ L(\psi)\r]^{-1}
   \l(\partial_t \psi +L(\psi)
   \frac{\delta{\cal F}[\psi]}{\delta\psi} \r)\, . 
\eeq
Computing the response fucntion using equ.~(\ref{Z_OM})
we obtain
\beq
\label{GR_OM_1}
 G_R(x_1,t_1;x_2,t_2) =  \frac{1}{2}\left\langle 
\psi(x_1,t_1)\l[
\partial_t \psi +L(\psi)
   \frac{\delta{\cal F}[\psi]}{\delta\psi}\r]
(x_2,t_2)\right\rangle \, . 
\eeq
In the following, we will denote $U[\psi]=L(\psi)
(\delta F[\psi])/(\delta\psi)$. The response function
is retarded and satisfies
\beq
\label{GR_OM_2}
  G_R(x_1,t_1;x_2,t_2) =  \frac{1}{2}\left\langle 
\psi(x_1,t_1)\l[ \partial_t \psi +
  U[\psi] \r] (x_2,t_2) \right\rangle 
 = 0 \, , \;\; t_1<t_2 \, . 
 \eeq
so that
\beq 
\label{psi-dot-U}
\left\langle 
\psi(x_1,t_1)\partial_t \psi(x_2,t) \right\rangle  
 - \left\langle \psi(x_1,t_1)  U[\psi] (x_2,t)
   \right\rangle  = 0 
\, , \;\; t_1<t \, . 
\eeq
We can now analzye the behavior of equ.~(\ref{psi-dot-U})
under the T-reversal symmetry defined in 
equ.~(\ref{T-rev-psi},\ref{T-rev-psit}). We find that 
the first term is even, whereas the second term is odd,
so that for $t_1>t$ the two terms in equ.~(\ref{GR_OM_2})
contribute equally. Comparing equ.~(\ref{GR_MSRJD}) and
equ.~(\ref{GR_OM_2}) then implies the 
fluctuation-dissipation relation
\beq
\label{FD-modB-4}
\left\langle \psi(x_1,t_1) \l[L(\psi)
  \tilde\psi\r](x_2,t_2)\right\rangle 
 = \Theta(t_2-t_1) \left\langle \psi(x_1,t_1)\dot\psi(x_2,t_2) 
 \right\rangle 
\eeq  
given in equ.~(\ref{FD-modB}). This relation can 
be generalized to more complicated response functions, 
and to model H. 




\bibliography{multi}
\end{document}